\documentclass[11pt]{article}

\renewcommand{\baselinestretch}{1.2}
  \renewcommand{\arraystretch}{1.1}
 \usepackage{amsmath}
\usepackage{amssymb}
\usepackage{graphicx}
\usepackage{ulem}
\usepackage{hyperref}

\begin{document}
 \title{Remarks on Quantum Modular Exponentiation
 and \\ Some Experimental Demonstrations of Shor's Algorithm}
  \author{Zhengjun Cao$^{1,*}$, \quad Zhenfu Cao,$^{2,3}$ \quad Lihua Liu$^{4}$}
  \footnotetext{ $^1$Department of Mathematics, Shanghai University, Shanghai,
  China. $^*$\,\textsf{caozhj@shu.edu.cn}\\
    $^2$Software Engineering Institute, East China Normal University, Shanghai,
  China.  \\
   $^3$Department of Computer Science and Engineering,  Shanghai Jiao Tong University,
  China.\\
     $^4$Department of Mathematics, Shanghai Maritime University,  Shanghai,
  China.     
   }

\date{}
\maketitle

\begin{quotation}
 \textbf{Abstract}.
  An efficient quantum modular exponentiation method is indispensible for Shor's factoring algorithm.  But we find that
all descriptions presented by Shor, Nielsen and Chuang, Markov and Saeedi, et al., are flawed.
 We also remark that some experimental demonstrations of Shor's algorithm are misleading, because they violate the necessary condition that the selected number $q=2^s$, where $s$ is the number of qubits used in the first register,  must satisfy
 $n^2 \leq  q < 2n^2$, where $n$ is the large number to be factored.

 \end{quotation}

 \section{Introduction}
It is well known that factoring an integer $n$ can be reduced to finding
the order of an integer $x$ with respect to the module $n$ (G. Miller \cite{M76}). The order is usually denoted by the notation $\mbox{ord}_n(x).$
So far, there is not a polynomial time algorithm run on classical computers which can be used to compute $\mbox{ord}_n(x)$.
In 1994, P. Shor \cite{S97} proposed the first quantum  algorithm which can  compute $\mbox{ord}_n(x)$ in polynomial time.
The factoring  algorithm  requires two quantum
registers. At the beginning of the algorithm, one has to
  find $q=2^s$ for some integer $s$ such that $n^2 \leq  q < 2n^2$, where $n$ is to be factored. The followed steps are:

  \begin{itemize}\parskip -2mm
\item[] \textit{Initialization}. Put register-1 in the following uniform superposition
$$ \frac 1{\sqrt q}\sum_{a=0}^{q-1}|a\rangle|0\rangle. $$

\item[] \textit{Computation}. Keep $a$ in
  register-1 and
  compute $x^a$ in register-2 for some randomly chosen integer $x$.  We then have the following state
$$ \frac 1{\sqrt q}\sum_{a=0}^{q-1}
|a\rangle|x^a\rangle. $$

\item[] \textit{Fourier transformation}.   Performing Fourier transform on register-1, we obtain the state
$$ \frac 1{q}\sum_{a=0}^{q-1}\sum_{c=0}^{q-1}
\mbox{exp}(2\pi iac/q)|c\rangle|x^a\rangle.
$$

\item[] \textit{Observation}.  It suffices  to
observe the first register.  The probability $p$ that the machine reaches the  state $|c, x^k\rangle$ is
$$  \left| \frac{1}{q}\, \sum_{a:\, x^a \equiv x^k} \mbox{exp}(2\pi i ac/q) \right|^2 $$
where $ 0\leq k< r=\mbox{ord}_n(x) $,  the sum is over all $a\, (0 \leq a < q)$  such that $x^a\equiv
x^k$.

\item[] \textit{Continued fraction expansion}.
If there is a $d$ such that $\frac{-r}2 \leq dq-rc\leq \frac r 2  $,
then the probability of seeing $|c, x^k\rangle$ is greater than $1/3r^2$.
Hence, we have
$$ \left|\frac c q-\frac d r \right|\leq \frac 1 {2q}\leq  \frac 1 {2n^2} <  \frac 1 {2r^2}.$$
\uwave{Since  $q\geq  n^2$,  we can round $c/q$ to obtain  $d/r$.} Thus  $r$ can be obtained.
\end{itemize}

 P. Shor has specified the operations for the process $ |0\rangle|0\rangle\rightarrow \frac 1{\sqrt q}\sum_{a=0}^{q-1}|a\rangle|0\rangle$, but not specified the operations  for
the process
$ \frac 1{\sqrt q}\sum_{a=0}^{q-1}|a\rangle|0\rangle\rightarrow \frac 1{\sqrt q}\sum_{a=0}^{q-1}
|a\rangle|x^a(\mbox{mod}\, n)\rangle.$
His original description specifies only the process  $(a, 1)\rightarrow(a, x^a\,\mbox{mod}\, n)$.
  Nielsen and Chuang in their book Ref.\cite{NC00} specify that
 $$  |a\rangle|y\rangle
   \rightarrow  |a\rangle U^{a_{t-1}2^{t-1}}\cdots U^{a_02^{0}}|y \rangle
 = |a\rangle|x^{a_{t-1}2^{t-1}} \times \cdots \times x^{a_0 2^0}y (\mbox{mod}\, n)\rangle
 = |a\rangle|x^{a}y (\mbox{mod}\, n)\rangle $$
where $a$'s binary representation is $a_{t-1}a_{t-2}\cdots a_0$, $U$ is the unitary operation such that $U|y\rangle\equiv |xy (\mbox{mod}\, n)\rangle$,  $y\in\{0, 1\}^{\ell}$,
 $\ell$ is the bit length of $n$.

 We find the Nielsen-Chuang quantum modular exponentiation method requires  $a $ unitary operations.
 Apparently, it is inappropriate for the process
 $$ \frac 1{\sqrt q}\sum_{a=0}^{q-1}|a\rangle|0\rangle \rightarrow \frac 1{\sqrt q}\sum_{a=0}^{q-1}
|a\rangle|x^a(\mbox{mod}\, n)\rangle $$
where $n^2 \leq  q < 2n^2$ and  $n$ is the large number to be factored, because the total amount of unitary operations required for this process  is $O(q^2)$,  not $O(\log n)$.
So far, there are few literatures to investigate the above mysterious process. In view of that $O(q^2)$ unitary operations can not be implemented in polynomial time,
we do not think that Shor's factoring algorithm is completely understandable and universally acceptable.

Since 2001, some teams have reported that they had successfully factored 15 into $3\times 5$ using Shor's algorithm. We shall have a close look at these experimental demonstrations and
remark that these demonstrations are misleading, because they violate the necessary condition that the selected number $q$ must satisfy
 $n^2 \leq  q < 2n^2$.

\section{Preliminaries}

A quantum analogue of a classical computer operates with quantum bits involving quantum states. The state of a quantum computer is described as a basis vector in a Hilbert space.
A qubit is a quantum state $|\Psi\rangle $ of the form
$$|\Psi\rangle= a|0\rangle+b|1\rangle, $$
where the amplitudes $a, b\in \mathbb{C}$ such that $|a|^2+|b|^2=1,$  $|0\rangle$ and $|1\rangle$ are basis vectors of the Hilbert space. Here, the \textit{ket} notation $|x\rangle$ means that $x$ is a quantum state.
 The state of a quantum system having $n$ qubits is a
point in a $2^n$-dimensional vector space. Given a state
$$\sum_{i=0}^{2^n-1} a_i|\chi_i\rangle, $$ where the amplitudes are complex numbers such that $\sum_{i=0}^{2^n-1}|a_i|^2=1$ and each
$|\chi_i\rangle$ is a basis vector of the Hilbert space, if the machine is measured with respect to this basis, the probability of seeing basis state $|\chi_i\rangle$ is $|a_i|^2$.

\textit{Two quantum mechanical systems are combined using the tensor product}.
For example,   a system of two qubits $|\Psi\rangle= a_1|0\rangle+a_2|1\rangle $ and $|\Phi \rangle= b_1|0\rangle+b_2|1\rangle $ can be written as
$$ |\Psi\rangle|\Phi \rangle= {a_1 \choose a_2}\otimes {b_1 \choose b_2}=
\left(\begin{array}{c}
   a_1b_1 \\
  a_1b_2 \\
  a_2b_1 \\
  a_2b_2 \\
 \end{array}
\right)
  $$
  We shall also use the shorthand notations  $|\Psi, \Phi \rangle$. We call a quantum  state having two or more components  \textit{entangled} state, if
it is  not a product state. According to the Copenhagen interpretation of quantum mechanics, measurement causes an instantaneous collapse of the wave function describing the quantum system into an eigenstate of the observable state that was measured.
If entangled, one object cannot be fully described without considering the other(s).

Operations on a qubit are described by $2\times 2$ unitary matrices. Of these, some of the most important are
 $$
X= \left[
\begin{array}{cc}
 0 & 1 \\
 1 & 0 \\
 \end{array}
 \right], \
 Y= \left[
\begin{array}{cc}
 0 & -i \\
 i & 0 \\
 \end{array}
 \right], \
  Z= \left[
\begin{array}{cc}
 1 & 0 \\
 0 & -1 \\
 \end{array}
 \right], \
  H=\frac 1 {\sqrt 2} \left[
\begin{array}{cc}
1 & 1 \\
1 & -1 \\
\end{array}
\right],
$$
where $H$ denotes the Hadamard gate. Clearly, $H|0\rangle= \frac 1 {\sqrt 2}(|0\rangle+|1\rangle). $

Operations on two qubits are described by $4\times 4$ unitary matrices. Of these,  the most important operation is the controlled-NOT, denoted by CNOT. The action
of CNOT is given by
$|c\rangle|t\rangle\rightarrow |c\rangle|c \oplus t\rangle$, where $\oplus $ denotes addition modulo 2. The matrix representation of CNOT is
$$
\left[
\begin{array}{cccc}
 1 & 0 & 0& 0\\
 0 & 1 &0 &0 \\
 0 & 0 & 0& 1\\
 0 & 0 & 1& 0
 \end{array}
 \right].$$
 Likewise, \uwave{operations on $\ell$ qubits are described by $2^{\ell}\times 2^{\ell}$ unitary matrices}.

 There is another method to describe linear operators performed on \textit{multiple qubits}. Suppose that $V$ and $W$ are vector spaces of dimension $2^{\mu}$ and $2^{\nu}$ (they describe quantum systems corresponding to $\mu$ and $\nu$ qubits, respectively). Suppose $|v\rangle$ and $|w\rangle$ are vectors in $V$ and $W$, and $A$ and $B$ are linear operators on $V$ and $W$, respectively.
Then we can define a linear operator $A\otimes B$ on $V\otimes W$ by the equation
$$(A\otimes B) (|v\rangle \otimes |w\rangle) \equiv A|v\rangle \otimes B |w\rangle.$$

\section{Remarks on quantum modular exponentiation method}

\subsection{The Shor's original description}

P. Shor has specified the operations for the process
$$ |0\rangle|0\rangle\rightarrow \frac 1{\sqrt q}\sum_{a=0}^{q-1}|a\rangle|0\rangle,$$
where $q=2^s$ for some positive integer $s$ such that $n^2 \leq  q < 2n^2$,  $n$ is to be factored. Notice that the first register consists of $s$ qubits.
He wrote: ``this step is relatively easy, since all it entails is putting each qubit in the first register into
the superposition $\frac 1{\sqrt 2}(|0\rangle+|1\rangle).$" (This can be done using the Hadamard gate $s$ times.)

Shor has not specified the operations  for
the process
$$ \frac 1{\sqrt q}\sum_{a=0}^{q-1}|a\rangle|0\rangle\rightarrow \frac 1{\sqrt q}\sum_{a=0}^{q-1}
|a\rangle|x^a(\mbox{mod}\, n)\rangle.$$
By the way, he has not specified how many qubits are required in the second register.
His original description specifies only the process  $(a, 1)\rightarrow(a, x^a\,\mbox{mod}\, n)$.
For convenience, we now relate it as follows.
\vspace*{3mm}

\fbox{\shortstack[l]{The technique  for computing $x^a \,(\mbox{mod}\,)$ is essentially the same as the classical method. \\ First, by repeated squaring we compute $x^{2^i}\,(\mbox{mod}\,)$ for all $i<l$. Then, to obtain $x^a \,(\mbox{mod}\,)$ \\
we multiply the powers $x^a \,(\mbox{mod}\,)$ where $2^i$ appears in the binary expansion of $a$. In our\\
 algorithm for factoring $n$, we only need to compute $x^a \,(\mbox{mod}\,)$ \underline{where $a$ is in a superposition}  \\
of states, but $x$ is some fixed integer. This makes things much easier, because we can use a \\
 reversible gate array where $a$ is treated as input, but where $x$ and $n$ are built into the \\
structure of the gate array. Thus, we can use the algorithm described by the following\\
  pseudocode;  here, \underline{$a_i$ represents the $i$th
bit of $a$ in binary}, where the bits are indexed from \\
 right to left and the rightmost bit of $a$ is $a_0$.\\
 \\
\hspace*{20mm}\textit{power}:=1\\
\hspace*{20mm}\textsf{for} $i=0$ \textsf{to} $l-1$\\
\hspace*{26mm}\textsf{if} $(a_i==1)$ \textsf{then} \\
\hspace*{32mm}\textit{power}:=\textit{power} $*$ $x^{2^i} \,(\mbox{mod}\, n)$\\
\hspace*{26mm}\textsf{endif}\\
\hspace*{20mm}\textsf{endfor}\\
\\
The variable $a$ is left unchanged by the code and $x^a \,(\mbox{mod}\,)$ is output as the variable \textit{power}.\\
 Thus, this code takes the pair of values $(a, 1)$ to $(a, x^a\,(\mbox{mod}\,))$.
}}\vspace*{3mm}

Remarks on the Shor's description:
\begin{itemize}
\item{} The description indicates only the conventional process $$(a, 1)\rightarrow (a, x^a\,\mbox{mod}\, n),$$ rather than the quantum process
$$|a\rangle |0\rangle\rightarrow |a\rangle |x^a\,\mbox{mod}\, n\rangle,$$ let alone the more complicated quantum process
$$ \frac 1{\sqrt q}\sum_{a=0}^{q-1}|a\rangle|0\rangle\rightarrow \frac 1{\sqrt q}\sum_{a=0}^{q-1}
|a\rangle|x^a(\mbox{mod}\, n)\rangle.$$

\item{} Since $a_i$ is required to compute $x^a(\mbox{mod}\, n) $ which  represents the $i$th bit of $a$ in binary,  one has to measure the superposition
$\frac 1{\sqrt q}\sum_{a=0}^{q-1}|a\rangle|0\rangle$ to obtain $a$. But it is impossible to practically compose pure states
$$|a\rangle|x^a(\mbox{mod}\, n)\rangle,\   a=0, 1, \cdots, q-1, $$
into the superposition $\frac 1{\sqrt q}\sum_{a=0}^{q-1}
|a\rangle|x^a(\mbox{mod}\, n)\rangle, $
because $q\geq n^2$  and $n$ is the large number to be factored.

\item{} Although it specifies the Hadamard gate  on each qubit in the first register, \uwave{it does not specify how many and what quantum gates or unitary operations are used on each qubit or a group of qubits in the second quantum register. }
 \end{itemize}

\subsection{The Nielsen-Chuang description}

 Nielsen and Chuang in their book Ref.\cite{NC00} specify that
 $$  |a\rangle|y\rangle
   \rightarrow  |a\rangle U^{a_{t-1}2^{t-1}}\cdots U^{a_02^{0}}|y \rangle
 = |a\rangle|x^{a_{t-1}2^{t-1}} \times \cdots \times x^{a_0 2^0}y (\mbox{mod}\, n)\rangle
 = |a\rangle|x^{a}y (\mbox{mod}\, n)\rangle $$
where $a$'s binary representation is $a_{t-1}a_{t-2}\cdots a_0$, $U$ is the unitary operation such that
$$U|y\rangle\equiv |xy (\mbox{mod}\, n)\rangle,$$  $y\in\{0, 1\}^{\ell}$,
 $\ell$ is the bit length of $n$.
They wrote:\vspace*{3mm}

\hspace*{5mm}\fbox{\shortstack[l]{
  Using the techniques of Section 3.2.5, it is now straightforward to construct a\\
   reversible circuit with
a $t$ bit register and an $\ell$ bit register which, when started\\
 in the state $(a, y)$ outputs $(a, x^ay(\mbox{mod}\, n))$,
using $O(\ell^3)$ gates, which can be \\
translated into a quantum circuit using $O(\ell^3)$ gates computing the transformation \\
$|a\rangle|y\rangle\rightarrow |a\rangle|x^ay (\mbox{mod}\, n)\rangle$.
}} \vspace*{3mm}

 Although they indicate that the classical circuit for the conventional process
$$(a, y)\stackrel{O(\ell^3) \ \mbox{classical gates}}{-------\longrightarrow} (a, x^ay(\mbox{mod}\, n))$$
can be translated into a quantum circuit for the quantum process
 $$|a\rangle|y\rangle \stackrel{O(\ell^3) \ \mbox{quantum gates}}{-------\longrightarrow} |a\rangle|x^ay (\mbox{mod}\, n)\rangle,$$
we now want to remark that the quantum circuit has to invoke $U$, the unitary operation, \underline{$a$ times}.  Thus, the wanted process
  $$ \frac 1{\sqrt q}\sum_{a=0}^{q-1}|a\rangle|0\rangle\rightarrow \frac 1{\sqrt q}\sum_{a=0}^{q-1}
|a\rangle|x^a(\mbox{mod}\, n)\rangle$$
has to invoke the unitary operation $1+2+\cdots +(q-1) \approx O(q^2)$ times,  if all terms $|a\rangle|0\rangle $, $a=0, \cdots, q-1$, are
processed one by one.  Even worse, the transformation for the process
  $$  |q-1\rangle|y\rangle
   \rightarrow  |a\rangle|x^{q-1}y (\mbox{mod}\, n)\rangle $$
has to invoke the unitary operation $q-1$ times according to the Nielsen-Chuang description. Clearly,
 \uwave{it can not be accomplished in polynomial time because $q$ is a large number}.

\subsection{The Markov-Saeedi quantum circuit}

In recent, Markov and Saeedi  \cite{MS12, MS13} have proposed a quantum circuit for modular exponentiation. We refer to the following Figure 1 for the outline of their circuit.

 \begin{minipage}{0.9\textwidth}
\hspace*{15mm}\includegraphics[angle=0,height=5cm,width=13cm]{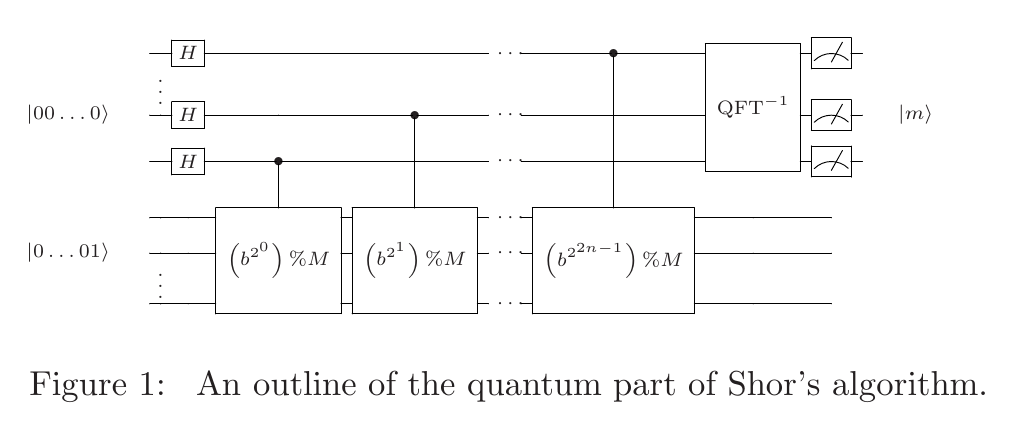}
 \end{minipage}
  \vspace*{-2mm}

The Markov-Saeedi quantum circuit for modular exponentiation is flawed, too.
   The unitary matrix corresponding to \fbox{$(b^{2^i}) \% M $ } for some integer $i$,  which is performed on \underline{all qubits} in the second quantum registers,  has a \underline{tremendous} dimension (not less than the modular $M$). To implement the operator practically, \uwave{one must decompose it into the tensor product of  some  linear operators with low dimension.} Regretfully, they had not specified these low dimension linear operators at all. Moreover,
   they had not specified the output of the operator \fbox{$(b^{2^0}) \% M $ }.
    We now want to ask:
\begin{itemize}
\item[] (1) what are the inputting  states of the unitary operator \fbox{$(b^{2^{2 n-1}}) \% M $ }?

\item[] (2) how to decompose the operator \fbox{$(b^{2^{2 n-1}}) \% M $ } into the tensor product of some low dimension linear operators?

\item[] (3) \uwave{how many executable unitary operators are required} in the quantum modular exponentiation process?
\end{itemize}
In our opinion, their proposed quantum circuit for modular exponentiation is incorrect and misleading.

\subsection{On Scott Aaronson's explanation}

 We have reported the flaw to some researchers including P. Shor himself, but only received a comment made by MIT professor Scott Aaronson. He  explained that (personal communication, 2014/09/02): \\ \vspace*{-1mm}

 \hspace*{5mm}\fbox{\shortstack[l]{
 The repeated squaring algorithm works (and works in polynomial time)\\
for any single $|a\rangle|0\rangle$, mapping it to $|a\rangle|x^a \,(\mbox{mod}\, n)\rangle$. But, because of the\\
 linearity of quantum
mechanics, this immediately implies that the algorithm\\
 must also work
for any superposition of $|a\rangle$'s, mapping $\sum_a |a\rangle$ to $\sum_a |a\rangle |x^a \,(\mbox{mod}\, n)\rangle$. }} \vspace*{3mm}\\
We do not think that his answer is convincing, because it is too vague to specify  \textit{how many and what quantum gates or unitary operations are used on each qubit or a group of qubits in the second quantum register}. Besides, according to the Nielsen-Chuang description, the process $$  |a\rangle|y\rangle
   \rightarrow  |a\rangle U^{a_{t-1}2^{t-1}}\cdots U^{a_02^{0}}|y \rangle
 = |a\rangle|x^{a_{t-1}2^{t-1}} \times \cdots \times x^{a_0 2^0}y (\mbox{mod}\, n)\rangle
 = |a\rangle|x^{a}y (\mbox{mod}\, n)\rangle $$
 depends on the binary representation of the exponent $a$.
 Which integer should be extracted in the superposition $ \frac 1{\sqrt q}\sum_{a=0}^{q-1}
|a\rangle|0\rangle$   for computing the wanted state  $ \frac 1{\sqrt q}\sum_{a=0}^{q-1}
|a\rangle|x^a(\mbox{mod}\, n)\rangle$? He did not pay more attentions to \textit{the difference between two linear operators performed on a pure state and a superposition}.

\section{It is difficult to modulate the wanted state in the second register}

We know the wanted superposition in the first register is modulated by the following procedure.
First, a Hadamard gate
$
H=\frac 1 {\sqrt 2} \left[
\begin{array}{cc}
1 & 1 \\
1 & -1 \\
\end{array}
\right]$  is performed on each qubit to obtain the $s$ intermediate states of $\frac 1 {\sqrt 2}(|0\rangle+|1\rangle)$. Second, combine all these states using the tensor product.
\begin{eqnarray*}
& & \frac 1{\sqrt 2}(|0\rangle+|1\rangle)\otimes\frac 1{\sqrt 2}(|0\rangle+|1\rangle)
= \frac 1{2}(|00\rangle+|01\rangle+|10\rangle+|11\rangle)\\
& & \frac 1{\sqrt 2}(|0\rangle+|1\rangle)\otimes\frac 1{\sqrt 2}(|0\rangle+|1\rangle)\otimes\frac 1{\sqrt 2}(|0\rangle+|1\rangle)\\
&=& \frac 1{2\sqrt 2}(|000\rangle+|001\rangle+|010\rangle+|011\rangle+|100\rangle+|101\rangle+|110\rangle+|111\rangle)\\
& & \vdots \\
& & \underbrace{\frac 1{\sqrt 2}(|0\rangle+|1\rangle)\otimes\cdots\otimes\frac 1{\sqrt 2}(|0\rangle+|1\rangle)}_{s\ \mbox{qubits}}
= \frac 1{\sqrt q}\sum_{a=0}^{q-1}
|a\rangle
\end{eqnarray*}
Note that the procedure works well because all those involved pure states are in  binary form.

We would like to stress that if two pure states are in decimal representations $|x\rangle, |x^2\rangle$,  then we can not directly combine them to obtain $|x^3\rangle$. Suppose that the binary strings for integers $x, x^2$ are $b_k\cdots b_0$, $b'_i\cdots b'_0$. We have
$$|x\rangle\otimes|x^2\rangle=|b_k\cdots b_0b'_i\cdots b'_0\rangle=|2^{i+1}x+x^2\rangle. $$
Thus,
$$\frac 1{\sqrt 2}(|1\rangle+|x\rangle)\otimes\frac 1{\sqrt 2}(|1\rangle+|x^2(\mbox{mod}\, n)\rangle)\otimes\cdots \otimes\frac 1{\sqrt 2}\left(|1\rangle+|x^{2^{s-1}}(\mbox{mod}\, n)\rangle\right)\neq \frac 1{\sqrt q}\sum_{a=0}^{q-1}
|x^a(\mbox{mod}\, n)\rangle,$$
where $q=2^s$,
although there is a corresponding conventional equation
$$(1+x)(1+x^2)(1+x^{2^2})\cdots (1+x^{2^{s-1}})= \sum_{a=0}^{q-1} x^a.  $$
\uwave{It seems that some people are confused by the above equation and simply take for granted that quantum modular exponentiation is in polynomial time.}

\section{On some experimental demonstrations of Shor's algorithm}

In 2001, it is reported that Shor's algorithm was demonstrated by a group at IBM, who factored 15 into $3\times 5$, using a quantum computer with \textit{7 qubits},
\textit{3 qubits for the first register} and 4 qubits for the second register (see Figure-2)  \cite{VS01}.

  In 2007, a group at University of Queensland  reported an experimental demonstration of a compiled version of Shor's algorithm. They factored 15 into $3\times 5$, using \textit{7 qubits} either, \textit{3 qubits for the first register} and 4 qubits for the second register (see Figure-3) \cite{LW07}.

  In 2007, a group at University of Science and Technology of China  reported another experimental demonstration of a complied version of Shor's algorithm.
They factored 15 into $3\times 5$ using \textit{6 qubits} only, \textit{2 qubits for the first register} and 4 qubits for the second register (see Figure-4) \cite{LB07}.

In 2012, a group at University of California, Santa Barbara,  reported a new experimental demonstration of a compiled version of Shor's algorithm. They factored 15 into $3\times 5$ using \textit{3 qubits} either, \textit{1 qubits for the first register} and 2 qubits for the second register (see Figure-5)  \cite{L12}.

\begin{center} \begin{tabular}{|l|c|c|}
  \hline
   Demonstrations  & qubits used in the first register  & qubits used in the second register  \\ \hline
  Figure 2, Ref.\cite{VS01}& 3 & 4 \\
  Figure 3,  Ref.\cite{LW07}& 3 & 4 \\
 Figure 4,   Ref.\cite{LB07}& 2 & 4 \\
  Figure 5,  Ref.\cite{L12}& 1 & 2\\
  \hline
\end{tabular}\end{center}\vspace*{4mm}

 \begin{minipage}{0.9\textwidth}
\hspace*{6mm}\includegraphics[angle=0,height=7cm,width=14cm]{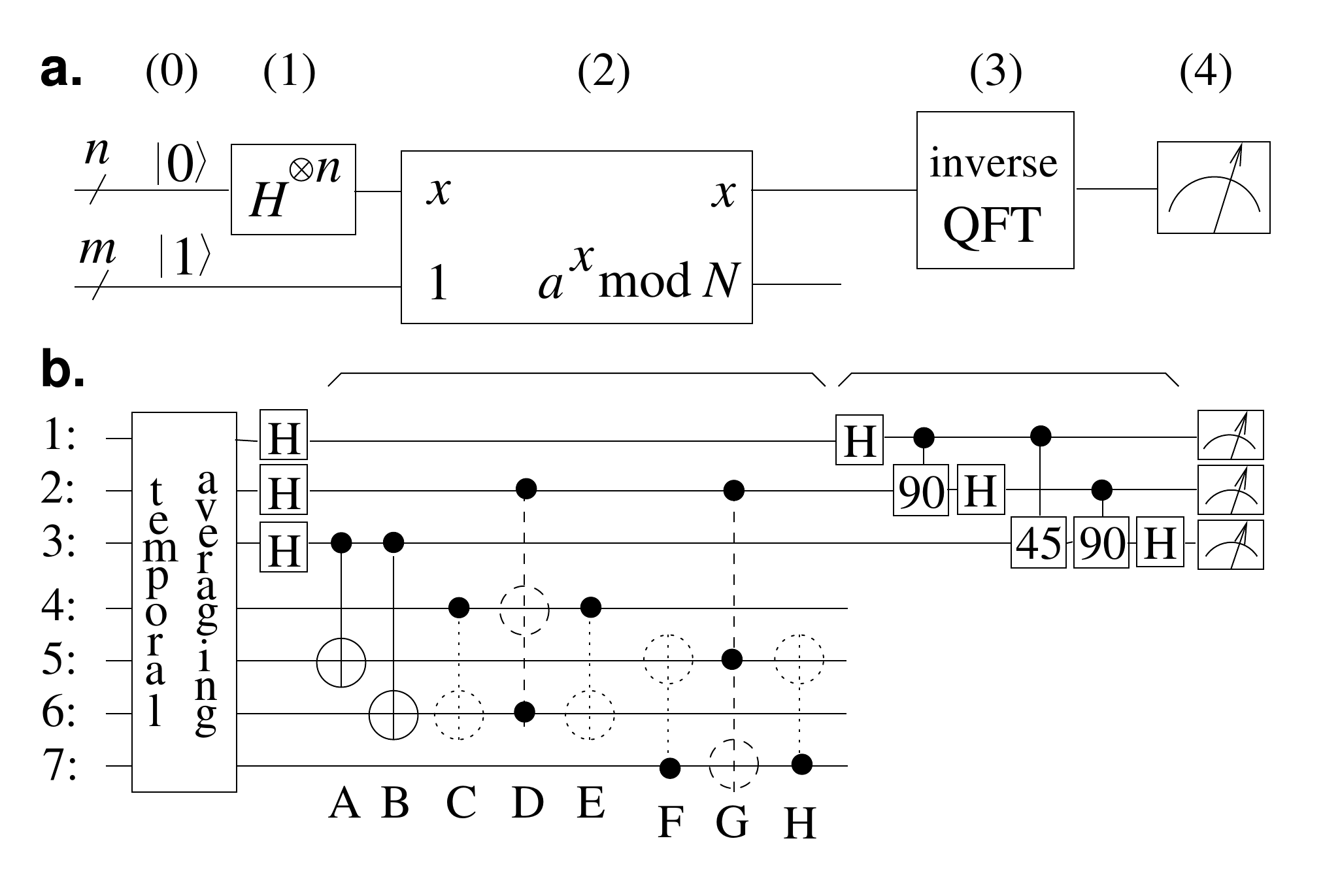}

\centerline{Figure 2:  Detailed quantum circuit for the case
$N = 15$ and $a = 7$.}
\end{minipage} \newpage

\begin{minipage}{0.9\textwidth}
 \hspace*{6mm}\includegraphics[angle=0,height=8cm,width=14cm]{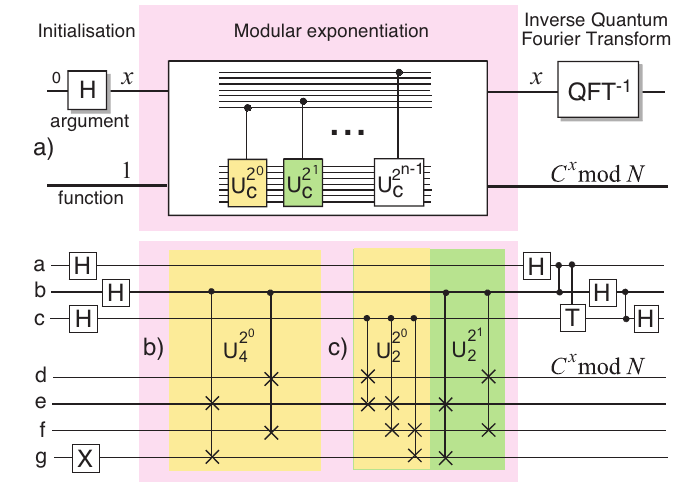}

\centerline{Figure 3:  Conceptual circuit for Shor's algorithm for number $N=15$ and co-prime $C=4$.}
 \end{minipage}
  \vspace*{4mm}

   \begin{minipage}{0.9\textwidth}
\hspace*{2mm}\includegraphics[angle=0,height=5cm,width=14cm]{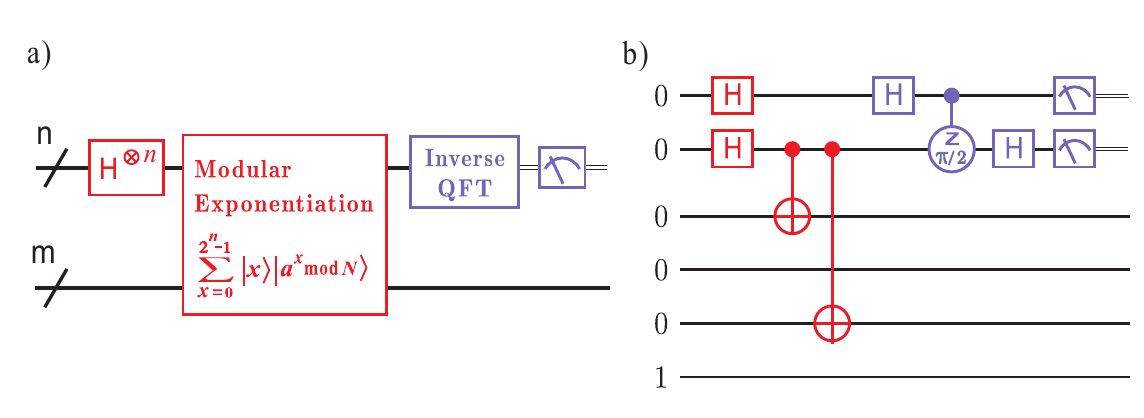}

\centerline{Figure 4:   Outline of quantum circuit for Shor's algorithm for $N = 15$ and $a = 11$.}
 \end{minipage}\vspace*{5mm}

     \begin{minipage}{0.9\textwidth}
 \hspace*{15mm}\includegraphics[angle=0,height=6.5cm,width=11cm]{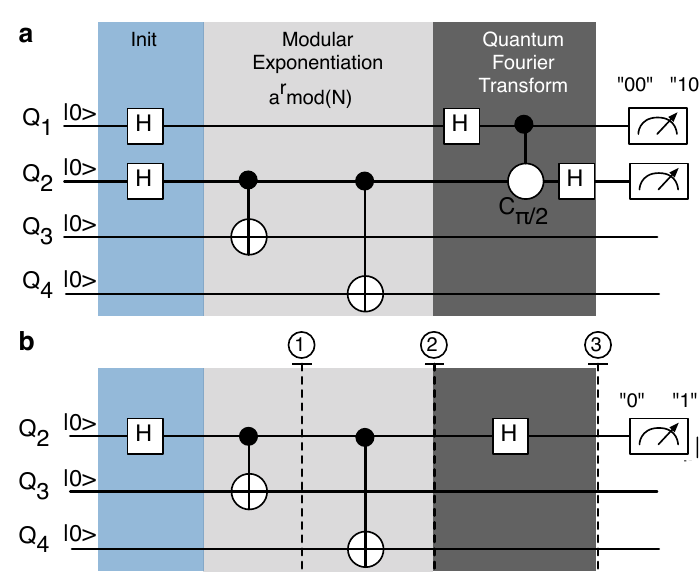}

\centerline{Figure 5:  A three-qubit compiled version of
Shor's algorithm to factor  $N = 15$.}
 \end{minipage}

 We now want to remark that:
  \begin{itemize}
 \item{} All these demonstrations are flawed because they violate the necessary condition that $15^2<2^8< 2 \times 15^2$, \uwave{which means 8 qubits should be used  in the first register.}  Obviously, the last step of continued fraction expansion in Shor's algorithm can not be accomplished if less qubits are used in the first register. It seems that these groups have misunderstood the necessary condition that $n^2\leq q<2n^2$  in Shor's algorithm.

\item{} In Figure 3, it directly denotes the output of the second register by $C^x \mod N $. Clearly, the authors confused the number $C^x \mod N $ with the state $|C^x \mod N\rangle $. By the way, the wanted  state in the second register is the superposition $\frac{1}{\sqrt 8}\sum_{x=0}^{7} |C^x \mod N\rangle $ instead of the pure state $|C^x \mod N\rangle $.

 \item{} In Figure 5, only 3 qubits are used. Clearly, the modular $15$ can not be represented by the 3 qubits.  In such case,
  \uwave{how to ensure that the modular is really involved in the computation}? In our opinion, the demonstration is unbelievable.
 \end{itemize}

    \section{Conclusion}
 Shor's factoring algorithm is interesting. But its subroutine for quantum modular exponentiation is not specified. We remark that both the Shor's original description
and the Nielsen-Chuang description for quantum modular exponentiation are flawed. They can  be used only for the \underline{pure state}
$|a\rangle|0\rangle$, not for the \underline{superposition}
$\frac 1{\sqrt q}\sum_{a=0}^{q-1}|a\rangle|0\rangle $. We also remark that some experimental demonstrations of Shor's algorithm are meaningless and misleading because they violate
a necessary condition for Shor's algorithm.

\textbf{Acknowledgements}. This work was supported by the National Natural Science Foundation of China (Grant Nos. 60970110, 60972034),
and the State Key Program of National Natural Science of China (Grant No. 61033014).

 \end{document}